\documentclass[12pt]{article}
\usepackage{amsmath,amsfonts}
\usepackage{algorithmic}
\usepackage{array}
\usepackage[caption=false,font=normalsize,labelfont=sf,textfont=sf]{subfig}
\usepackage{textcomp}
\usepackage{stfloats}
\usepackage{url}
\usepackage{verbatim}
\usepackage{graphicx}
\def\BibTeX{{\rm B\kern-.05em{\sc i\kern-.025em b}\kern-.08em
    T\kern-.1667em\lower.7ex\hbox{E}\kern-.125emX}}
\usepackage{balance}
\usepackage{amssymb}
\usepackage[sort]{cite}
\usepackage{setspace}
\usepackage[margin=1in]{geometry}
\usepackage{setspace}
\begin{document}
\title{Capacity Achieving Design for Hybrid Beamforming in Millimeter Wave Massive MIMO Systems}
\author{Rohollah Vahdani, and S. Mohammad Razavizadeh
\thanks{\copyright~2024 IEEE. Personal use of this material is permitted. However, permission to use this material for any other purposes must be obtained from the IEEE by sending a request to pubs-permissions@ieee.org.}
}
\date{}
\maketitle

\begin{abstract}
Hybrid  digital and analog beamforming is a highly effective technique for implementing beamforming methods in millimeter wave (mmWave) systems. It provides a viable solution to replace the complex fully digital beamforming techniques. However, the current design of precoding and combining matrices in hybrid beamforming solely relies on the channel information, neglecting the crucial consideration of the structure of covariance matrices of the transmit signals.
In this paper, we present a novel approach for the joint design of hybrid beamforming matrices at the transmitter and receiver. This approach is centered around the optimization of the covariance matrix of the transmitted signals. Our goal is to maximize the downlink sum rate capacity of the system by achieving an optimal design of the transmit covariance matrix. We tackle the non-convex nature of this problem by leveraging the dual relationship between the broadcast channel (BC) and the multiple access channel (MAC).
Through extensive simulations in various scenarios, including point-to-point multi-input multi-output (MIMO), multi-user (MU) multi-input single-output (MISO), and MU-MIMO, we demonstrate the superiority of our proposed method over traditional designs. These results highlight the effectiveness and versatility of  our approach in optimizing beamforming for mmWave systems.
\end{abstract}

\section{Introduction} \label{sec.intro}

Millimeter-wave (mmWave) technology is a promising technique in next-generation cellular systems to provide high bandwidth demands. One of the proposed methods for combating high path loss in mmWave frequencies is using large-scale antenna arrays or massive multi-input multi-output (MIMO) systems. In the conventional fully digital beamforming methods, each antenna element needs to be connected to a separate radio-frequency (RF) chain, which is not applicable in massive MIMO scenarios due to space and power consumption limitations. To solve this problem, hybrid digital and analog beamforming schemes have been proposed and widely investigated in recent years \cite{r7,r9,r12,r13}. In this technique, digital and analog beamforming matrices are jointly designed to reduce the number of RF chains \cite{r14}. It has been shown that the hybrid structure can achieve the performance of fully digital beamforming techniques \cite{r7}.
The main problem in the hybrid beamforming methods is finding the optimal precoders and combiners \cite{r12, r22, r15, r16, r17, r18, r19}. To optimize precoders and combiners, different criteria have been proposed, with sum rate maximization being the most common target \cite{r15, r16, r17, r18, r19}.
The authors in \cite{r15} have investigated a joint analog and digital precoder design at the base station (BS) transmitter with the aim of maximizing sum rate, utilizing the idea of singular value decomposition (SVD). However, the resulting non-convex sum rate maximization problem is difficult to solve. To address this, they have proposed methods to decompose the problem into a series of sub-problems.
In \cite{r16}, the sum rate capacity of a multi-user MIMO (MU-MIMO) system with a hybrid structure at the BS has been investigated, and partial swarm optimization (PSO) has been adopted for joint design of the hybrid digital and analog precoding.
In \cite{r17}, a multi-user multi-input single-output (MU-MISO) system with hybrid precoding at the BS (without hybrid structure at the user side) is considered. The authors have derived an asymptotic expression for the non-convex sum rate maximization problem in the downlink and proposed an algorithm based on semi-definite relaxation, which leads to suboptimal solutions.
The authors of \cite{r18} have designed analog and digital beamforming matrices separately. They used the multi-beam selection method to derive analog beamformers, while maximum-ratio combining (MRC) is used for digital beamforming \cite{r18}.
Additionally, the authors in \cite{r19} have attempted to find the optimal precoders and combiners using the minimum mean square error (MMSE) criterion. However, this method is not optimal from a sum rate capacity standpoint.
\\
Authors of \cite{rnew1} and \cite{rnew2} focused on design of hybrid transmit precoding for wideband multiuser mmWave MIMO systems, specifically in the presence of beam squint. They demonstrated that in massive MIMO systems operating in wideband millimeter wave frequencies, the direction of the beam in the spatial domain is dependent on the central frequencies. This is because digital beamforming is frequency-selective and varies across different ranges of central frequency. They proposed several full-CSI based hybrid transmit precoding schemes from a new perspective by projecting all frequencies to the central frequency and then constructing common analog beamforming matrices for all subcarriers. They utilized the long-term channel's covariance matrix and the angle of departure (AoD) information and constructed the analog transmit precoding matrix with infinite angular resolution and then proposed a non-uniformly spaced quantization codebook based analog precoding with finite angular resolution. \cite{rnew3} developed a fast-converging hybrid precoding algorithm and the authors of \cite{rnew4} proposed three dynamic hybrid beamforming schemes
to improve spectral efficiency in multiuser orthogonal frequency-division multiplexing (OFDM) systems. Additionally, near optimal sum-rate maximization schemes with finite resolution analog precoders and combiners were discussed in \cite{rnew5}. They designed the coefficients for finite resolution analog precoders and combiners as a lattice decoding problem, where low-complexity lattice decoders and convex solvers were employed to yield near optimal solutions. Furthermore, some studies on holographic MIMO systems have been conducted \cite{rnew6,rnew7}, notably one that formulates a joint optimization problem for Stacked intelligent metasurfaces (SIM) phase shifts and the transmitted signal's covariance matrix,
achieving comparable rates to existing benchmarks with fewer iterations\cite{rnew7}.

\begin{table*}[ht]
\centering
\caption{Comparison of the proposed hybrid beamforming design with other MIMO technologies}
    \begin{tabular}{ | c | c | c | c | c |}
    \hline
    MIMO technologies & Number of RF chains & Hardware cost & Sum-rate & Complexity\\ \hline
    Fully digital & Large & High & High & Very high \\ \hline
    Hybrid MMSE \cite{r10} & Moderate & High & High & High \\ \hline
    Hybrid Foad design \cite{r7} & Moderate & Moderate & Moderate & High \\ \hline
    Hybrid sequential-ZF \cite{r22} & High & High & Low & High \\ \hline
    Proposed hybrid digital-analog  & Small & Low & High & Low \\
    \hline
    \end{tabular}
\end{table*}

In almost all references on the design of hybrid beamforming, the authors have not considered the impact of the transmit covariance matrix, which is a key parameter in determining the capacity region of Gaussian interference channels. In fact, most papers assume Gaussian signaling with an identity covariance matrix in the broadcast channel (BC) and multiple-access channel (MAC), which is not optimal from an information-theoretic perspective.
The impact of the input covariance matrix on finding the capacity region of MU-MIMO systems has been widely studied in the literature, as seen in \cite{r27} and \cite{r28}. Furthermore, it has been shown that there is a duality relationship between the capacity region of the BC (downlink) and MAC (uplink). It is worth mentioning that in this paper, the link of the channel between BS and the users (BC) is called downlink and the inverse path (MAC) is called uplink because the system is considered from the user's point of view. This duality is very useful from an analytical standpoint because in the BC, the dirty paper region leads to non-concave rate functions of the individual users' covariance matrices, whereas the rate functions in the MAC are concave and can be mutually replaced \cite{r27}. In other words, the optimal covariance matrices in the MAC can be found using standard, low-complexity convex optimization techniques. Then, the optimal covariance matrices for the BC can be derived using the MAC-BC transformations given in \cite{r28}. Consequently, all rates achievable in the dual MIMO MAC are also achievable in the MIMO BC with the same power budgets, and vice versa, assuming the reciprocity between uplink and downlink channels \cite{r28}. This idea can be used in the context of hybrid beamforming and for the problem of precoding and decoding matrix design.

This paper addresses hybrid beamforming design in massive MIMO mmWave networks,
focusing on maximizing sum rate capacities in both downlink and uplink using the MAC-BC duality theorem. It formulates a non-convex sum rate maximization problem that involves individual transmit covariance and hybrid beamforming matrices, which is transformed into a more solvable dual MAC problem. An alternating algorithm will be proposed to optimize variables like transmit covariance matrices and hybrid precoders. Simulation shows that this method outperforms existing approaches, achieving sum rates comparable to fully-digital beamforming in high signal-to-noise ratios (SNRs), and surpassing other hybrid designs in lower SNR conditions across various scenarios.

The main contributions of this paper are summarized as follows:

\begin{itemize}
  \item We analysed the sum rate of the downlink and uplink in the hybrid mmWave massive MIMO systems considering different optimizing parameters (the user's covariance matrices, analog and digital beamforming/combining matrices). This will conclude better performance in comparison with prior works which employs identity covariance matrices for all users.
  \item We study the MAC/BC sum-rate maximization problem. and propose a heuristic algorithm to find the optimum solutions using MAC-BC duality theorem. The sum-rate maximization problem in the BC is non-concave, thus we transfer this problem into the dual MAC which the problem can be solved using standard, low complexity convex optimization techniques. Then, the optimal covariance matrices for the BC can be derived using the MAC-BC transformations.
  \item The simulation results demonstrate that our scheme outperforms existing designs, with significantly lower complexity.
\end{itemize}
Explicitly, we contrast our proposed design and the relevant technologies and concepts in Table I. The remainder of the paper is organized as follows:
In Section II, the channel and system models are described.
In Section III, an overview of the capacity regions in the MAC and BC is presented, along with the derivation of the sum rate optimization problems.
In Sections IV and V, the optimum hybrid beamforming and individual user's covariance matrices that maximize the sum rate are derived.
Simulation results and concluding remarks are presented in Sections VI and VII, respectively.

Notations: This paper uses upper-case bold face letters for matrices, and lower-case bold face letters for vectors. $(.)^{T}$, $(.)^{H}$ and $(.)^{\dagger}$ are used for transpose, Hermitian transpose and pseudo-inverse, respectively. Also $tr(.)$ denotes trace of a matrix and $\otimes$ is used for Kronecker product. $\textbf{I}_N$ is the identity matrix of size $N\times{N}$. Finally, $\mathbb{C}^{n\times{m}}$ is used to denote the complex matrix space with size $n\times{m}$.

\begin{figure*}[t!]\label{fig1}
  \centering
  \includegraphics[width=\textwidth]{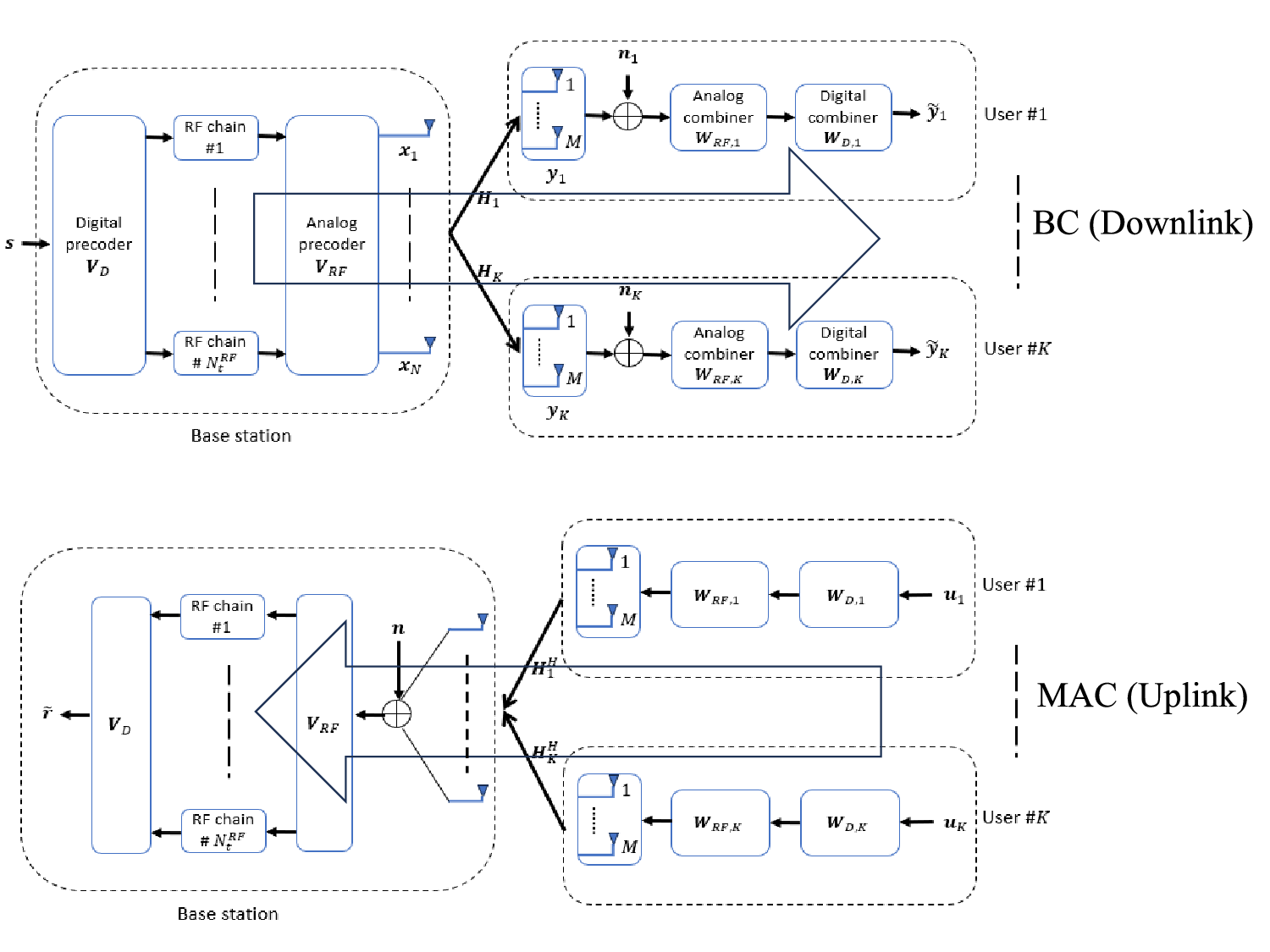}
  \caption{System model of the multi-user hybrid MIMO BC and MAC channels}
\end{figure*}

\section{System Model} \label{sec.sysmodel}

\subsection{Signal and System model} \label{signal}

In our system model, we consider a cellular-type BC in which a BS with $N$ antennas and $N^{RF}_{t}$ transmit RF chains serves $K$ user equipment, each with $M$ antennas as shown in Fig. 1. Both the BS and UEs use hybrid digital and analog structures. Here, we will focus on the BC or equivalently the downlink channel. Then, we will use the dual uplink channel, too. In the BS, the transmitter first precodes $N_{s}$ data symbols $\textbf{s}$ using low dimensional $N^{RF}_{t}\times{N_{s}}$ digital precoder $\textbf{V}_{D}$ followed by a $N\times{N^{RF}_{t}}$ RF precoder $\textbf{V}_{RF}$ to generate the final transmitted signal. Thus, the overall hybrid precoder can be expressed by $\textbf{V}=\textbf{V}_{RF}\textbf{V}_{D}$ with size $N\times{N_{s}}$. The digital precoder contains different digital precoder coefficients corresponding to different users each with size $N^{RF}_{t}\times{d}$. In other words, $\textbf{V}_{D}=[\textbf{V}^{1}_{D},\textbf{V}^{2}_{D},...,\textbf{V}^{K}_{D}]$ where $\textbf{V}^{i}_{D}$ is the digital precoder matrix corresponding to the $i$-th user. Then, the transmitted signal from the BS can be written as
\begin{equation} \label{eq.1}
  \textbf{x}=\textbf{V}\textbf{s}=\textbf{V}_{RF}\textbf{V}_{D}\textbf{s}=\textbf{V}\big(\textbf{s}_{1}+\textbf{s}_{2}+\ldots+\textbf{s}_{K}\big),
\end{equation}
where $\textbf{s}_{k}$ is the $N_{s}\times{1}$ individual transmitted stream corresponding to the $k$-th user. In the user side, the $M\times{1}$ received signal vector can be expressed as
\begin{equation} \label{eq.2}
\textbf{y}_{k}=\textbf{H}_{k}\textbf{x}+\textbf{n}_{k}=\textbf{H}_{k}\textbf{V}\textbf{s}_{k}+\textbf{H}_{k}\textbf{V}\sum_{l\neq{k}}{\textbf{s}_{l}}+\textbf{n}_{k},
\end{equation}
where $\textbf{H}_{k}\in{\mathbb{C}^{M\times{N}}}$ is the channel coefficient matrix between the BS and the $k$-th user and $\textbf{n}_{k}$ is the additive white Gaussian noise that each entry has zero mean and unit variance. The user $k$ processes the received signal using an analog RF combiner, $\textbf{W}_{RF,k}\in{\mathbb{C}^{M\times{N^{RF}_{r}}}}$. Then by employing a low dimensional digital combiner, $\textbf{W}_{D,k}\in{\mathbb{C}^{N^{RF}_{r}\times{N_{s}}}}$, the receiver obtains the final processed signal as
\begin{equation} \label{eq.3}
\tilde{\textbf{y}}_{k}=\textbf{W}^{H}_{k}\textbf{y}_{k}=\textbf{W}^{H}_{k}\textbf{H}_{k}\textbf{V}\textbf{s}_{k}+\textbf{W}^{H}_{k}\textbf{H}_{k}\textbf{V}\sum_{l\neq{k}}{\textbf{s}_{l}}+\textbf{W}^{H}_{k}\textbf{n}_{k},
\end{equation}
in which $\textbf{W}_{k}=\textbf{W}_{RF,k}\textbf{W}_{D,k}$ is the overall hybrid combiner for the $k$-th user.

Assuming the reciprocity between uplink and downlink channels, the uplink channel can be represented by $\textbf{H}^{H}_{k}$. In the MAC, the $N_{s}\times{1}$ processed signal at the BS can be expressed as
\begin{equation} \label{eq.4}
\tilde{\textbf{r}}=\textbf{V}^{H}\bigg(\sum_{j=1}^{K}{\textbf{H}^{H}_{j}\textbf{W}_{j}\textbf{u}_{j}}\bigg)+\textbf{V}^{H}\textbf{n},
\end{equation}
where $\textbf{u}_{j}$ is the $N_{s}\times{1}$ transmitted vector of streams in the MAC from the $j$-th user and $\textbf{n}$ is the additive white Gaussian noise vector in the BS.

\subsection{Millimeter Wave MIMO Channel Model } \label{mmWave channel}
Due to the high frequency of mmWave signals, the channel can be represented by multiple path that each directed from a specific antenna element in the transmitter to an antenna element in the receiver. Each path experience a delay proportional to the antenna locations and a complex gain. These paths can be concatenated into a vector called steering vectors of the transmitter and receiver, respectively. For an $N$-element uniform linear array (ULA), the steering vector is given by \cite{r2}
\begin{equation} \label{eq.5}
\textbf{a}(\theta_{k,l})=\bigg[1,e^{j2\pi\frac{d_{a}}{\lambda}\sin{(\theta_{k,l})}},\ldots,e^{j2\pi\frac{d_{a}}{\lambda}(N-1)\sin{(\theta_{k,l})}}\bigg]^{T},
\end{equation}
where  $\theta_{k,l}$ is the angle of arrival (or departure) of the transmitted signal from $k$-th antenna element in the $l$-th scattering path which is a random variable with uniform distribution $\theta_{k,l}\in{[0,2\pi]}$, $d_{a}$ is the space between each antenna elements and $\lambda$ is the wavelength. For simplicity and without loss of generality, it is assumed that $d_{a}=\lambda/2$. Finally, the mmWave channel between the BS to the $k$-th user can be represented as \cite{r2}
\begin{equation} \label{eq.6}
\textbf{H}_{k}=\sum_{l=1}^{L}{\alpha_{k,l}\textbf{a}_{R}(\theta_{k,l})\textbf{a}^{H}_{T}(\theta_{k,l})},
\end{equation}
where $\textbf{a}_{R}(\theta_{k,l})$ and $\textbf{a}_{T}(\theta_{k,l})$ are the transmit and receive steering vectors, respectively. $\alpha_{k,l}$ is the zero mean and unit variance complex i.i.d. Gaussian random gain of the $l$-th path to the $k$-th antenna element. It must be noted that the steering vectors can be generalized to a 3D model by considering of both horizontal (azimuth) angle $\theta$ and vertical (elevation) angle $\phi$. Then each steering vector can be represented by $\textbf{a}(\theta,\phi)=\textbf{a}_{H}(\theta)\bigotimes{\textbf{a}_{V}(\phi)}$. The system model can be extended to the wideband by considering the realization of the channel matrix in each subcarrier, separately.

In next section, the sum rate capacity region of the multi-user MAC and BC in hybrid mode are derived. Then, the duality theorem is presented to achieve the same sum rate regions in both scenarios and the required conditions are determined.

\section{Capacity region and sum rate expressions of hybrid beamforming in MAC and BC}\label{secIII}

In the MIMO BC, dirty paper coding (DPC) can be applied at the transmitter to achieve the capacity region of the MU-MIMO interference channel \cite{r14,r15}. In other words, the DPC is the capacity achieving strategy for the sum rate of the MIMO BC. Assume that the user’s priority is set such that $\mu_{1}\geq\mu_{2}\geq\ldots\geq\mu_{K}$ where $\mu_{j}$ is the priority of the $j$-th user. In the DPC, the transmitter chooses a codeword for the $j$-th receiver with a full knowledge of the codewords intended for receiver $j+1$ to $K$.  Thus the $j$-th receiver can cancel the interference of user $j+1$ to $K$. Notice that within this way, the receiver 1 which has the highest priority will not see any interference from other users. Consequently, the rate of the $j$-th user in the hybrid BC can be expressed as
\begin{equation} \label{eq.7}
\begin{split}
R^{BC}_{j}&=\frac{1}{2}\log{\frac{\big|\textbf{W}^{H}_{j}\textbf{W}_{j}+\textbf{W}^{H}_{j}\textbf{H}_{j}\textbf{V}(\sum_{l=1}^{j}{\textbf{D}_{l}})\textbf{V}^{H}\textbf{H}^{H}_{j}\textbf{W}_{j}\big|}{\big|\textbf{W}^{H}_{j}\textbf{W}_{j}+\textbf{W}^{H}_{j}\textbf{H}_{j}\textbf{V}(\sum_{l=1}^{j-1}{\textbf{D}_{l}})\textbf{V}^{H}\textbf{H}^{H}_{j}\textbf{W}_{j}\big|}}\\
&=\frac{1}{2}\log{\big|\textbf{I}_{N_{s}}+\textbf{A}^{-1}_{j}\textbf{W}^{H}_{j}\textbf{H}_{j}\textbf{V}\textbf{D}_{j}\textbf{V}^{H}\textbf{H}^{H}_{j}\textbf{W}_{j}\big|},
\end{split}
\end{equation}
where $\textbf{D}_{j}$ is the $N_{s}\times{N_{s}}$ covariance matrix of the transmitted signal related to the $j$-th user and $\textbf{A}_{j}$ is the $N_{s}\times{N_{s}}$ covariance matrix of the interference plus noise terms in the $j$-th user in the BC that can be derived as
\begin{equation} \label{eq.8}
\textbf{A}_{j}=\textbf{W}^{H}_{j}\textbf{W}_{j}+\textbf{W}^{H}_{j}\textbf{H}_{j}\textbf{V}(\sum_{l=1}^{j-1}{\textbf{D}_{l}})\textbf{V}^{H}\textbf{H}^{H}_{j}\textbf{W}_{j},
\end{equation}
It must be noted that in the BC (as shown in (1)), the transmitted signal is the sum of the transmitted signals of each user and as the data streams of users are independent, the overall covariance matrix of the transmitted signal is the sum of all individual covariance matrices. Moreover, to meet the power constraint, we should have $tr(\textbf{V}\textbf{D}_{j}\textbf{V}^{H})=P_{j}$. It can be interpreted from (3) and (7) that the effective channel corresponding to the $j$-th user in the BC is equal to $\textbf{W}^{H}_{j}\textbf{H}_{j} \textbf{V}$.
One of the important features of the rate equations in (7) related to the dirty paper strategy is that the rate equations are neither concave nor convex functions of the covariance matrices. This makes finding the rate region very difficult because the entire space of covariance matrices which meet the power constraint must be searched over.
In the dual MAC, the priority of the users are in reverse order and the successive decoding can be applied to achieve to the capacity region. Therefore, the individual rate for user $j$ in the multi-user hybrid MAC can be expressed as
\begin{equation} \label{eq.9}
\begin{split}
R^{MAC}_{j}&=\frac{1}{2}\log{\frac{\big|\textbf{V}^{H}\textbf{V}+\textbf{V}^{H}\big(\sum_{k=j}^{K}{\textbf{H}^{H}_{j}\textbf{W}_{k}\textbf{Q}_{k}\textbf{W}^{H}_{k}\textbf{H}_{j}}\big)\textbf{V}\big|}{\big|\textbf{V}^{H}\textbf{V}+\textbf{V}^{H}\big(\sum_{k=j+1}^{K}{\textbf{H}^{H}_{j}\textbf{W}_{k}\textbf{Q}_{k}\textbf{W}^{H}_{k}\textbf{H}_{j}}\big)\textbf{V}\big|}}\\
&=\frac{1}{2}\log{\big|\textbf{I}_{N_{s}}+\textbf{B}^{-1}_{j}\textbf{V}^{H}\textbf{H}^{H}_{j}\textbf{W}_{j}\textbf{Q}_{j}\textbf{W}^{H}_{j}\textbf{H}_{j}\textbf{V}\big|},
\end{split}
\end{equation}
where $\textbf{Q}_{j}$ is the $N_{s}\times{N_{s}}$ transmit covariance matrix of the $j$-th user in the MAC channel ($\textbf{Q}_{j}=E[\textbf{u}_{j}\textbf{u}^{H}_{j}]$) and  $\textbf{B}_{j}$ is the $N_{s}\times{N_{s}}$ covariance matrix of the interference plus noise term related to the $j$-th user in the MAC that can be derived as
\begin{equation} \label{eq.10}
\textbf{B}_{j}=\textbf{V}^{H}\textbf{V}+\textbf{V}^{H}\big(\sum_{k=j+1}^{K}{\textbf{H}^{H}_{k}\textbf{W}_{k}\textbf{Q}_{k}\textbf{W}^{H}_{k}\textbf{H}_{k}}\big)\textbf{V},
\end{equation}
It can be interpreted from (4) and (9) that the effective channel corresponding to the $k$-th user in the MAC is equal to $\textbf{V}^{H}\textbf{H}^{H}_{k}\textbf{W}_{k}$, which is still the reciprocity of the effective channel of the MIMO BC. Then, the sum rate in hybrid MAC can be achieved as \cite{r29}
\begin{equation} \label{eq.11}
\begin{split}
&R^{MAC}=\sum_{j=1}^{K}{R^{MAC}_{j}}\\
&=\frac{1}{2}\log{\big|\textbf{V}^{H}\textbf{V}+\textbf{V}^{H}\big(\sum_{k=1}^{K}{\textbf{H}^{H}_{k}\textbf{W}_{k}\textbf{Q}_{k}\textbf{W}^{H}_{k}\textbf{H}_{k}}\big)\textbf{V}\big|}\\
&-\frac{1}{2}\log{|\textbf{V}^{H}\textbf{V}|},
\end{split}
\end{equation}
where $\textbf{V}^{H}\textbf{V}$ can be interpreted as the covariance of the noise term which is multiplied by $\textbf{V}$, thus its covariance matrix will be changed from $\textbf{I}$ to $\textbf{V}^{H}\textbf{V}$. In other words, if the matrix $\textbf{V}$ be semi-unitary, the second term in (11) will be omitted and the noise will not be colored. The precoding and combining matrices can be found by different criteria such as MMSE, ZF, or with the aim of maximizing the sum rate. The MIMO MAC rates are concave functions of the transmit covariance matrices, $\textbf{Q}_{k}$. This implies that the boundary points of the capacity region can be found by a standard convex programs to characterize the optimal covariance matrices $(\textbf{Q}_{1},\ldots,\textbf{Q}_{K})$ \cite{r28}.

The sum rate maximization problem in the dual BC and MAC can be written respectively as
\begin{subequations}\label{eq.12}
\begin{eqnarray}
\begin{aligned}
\max_{\textbf{V},\textbf{W}_{k},\textbf{D}_{k}} \quad & \bigg(\sum_{k=1}^{K}{\frac{1}{2}\log{\big|\textbf{I}_{N_{s}}+\textbf{A}^{-1}_{k}\textbf{W}^{H}_{k}\textbf{H}_{k}\textbf{V}\textbf{D}_{k}\textbf{V}^{H}\textbf{H}^{H}_{k}\textbf{W}_{k}\big|}}\bigg)\\
\textrm{s.t.} \quad & \begin{cases}
\sum_{k=1}^{K}{tr(\textbf{V}\textbf{D}_{k}\textbf{V}^{H})}\leq{P},\\
\big|\textbf{V}_{RF}(i,j)\big|=1,
\end{cases}
\end{aligned}
\end{eqnarray}
\begin{eqnarray}
\begin{aligned}
\max_{\textbf{V},\textbf{W}_{k},\textbf{Q}_{k}} \quad \bigg(&\frac{1}{2}\log{\big|\textbf{V}^{H}\textbf{V}+\textbf{V}^{H}\big(\sum_{k=1}^{K}{\textbf{H}^{H}_{k}\textbf{W}_{k}\textbf{Q}_{k}\textbf{W}^{H}_{k}\textbf{H}_{k}}\big)\textbf{V}\big|}\\
&-\frac{1}{2}\log{\big|\textbf{V}^{H}\textbf{V}\big|}\bigg)\\
\textrm{s.t.} \quad & \begin{cases}
\sum_{k=1}^{K}{tr(\textbf{W}_{k}\textbf{Q}_{k}\textbf{W}^{H}_{k})}\leq{P},\\
\big|\textbf{W}_{RF,k}(i,j)\big|=1,
\end{cases}
\end{aligned}
\end{eqnarray}
\end{subequations}
Based on the MAC-BC duality theorem (refer to the theorem 2 in \cite{r27}), the dirty paper region of the MIMO BC with power constraint $P$ is equal to the capacity region of the dual MIMO MAC with sum power constraint  $\sum_{k=1}^{K}{P_{k}}=P$. Thus, although BC maximization problem as a function of the individual covariance matrices is neither concave nor convex, instead we can solve the convex optimization problem of the MAC and then transform the optimum covariance matrices in the MAC to BC. By this way, the computational complexity will be reduced remarkably. Assume that the optimum values of the covariance matrices and the hybrid precoding and combiners are found in the MAC. As the effective channels of the BC ($\tilde{\textbf{H}}_{k}=\textbf{W}^{H}_{k}\textbf{H}_{k}\textbf{V}$) and MAC ($\tilde{\textbf{H}}^{H}_{k}=\textbf{V}^{H}\textbf{H}^{H}_{k}\textbf{W}_{k}$) have reciprocity, the MAC-BC duality theorem can be used if the transmit power constraint of both BC and MAC remain unchanged. To fulfill this condition, $\textbf{V}$ and $\textbf{W}_{k}$ should satisfy below constraints
\begin{subequations}\label{eq.13}
\begin{eqnarray}
\begin{aligned}
\text{BC:}\quad \sum_{k=1}^{K}{tr(\textbf{V}\textbf{D}_{k}\textbf{V}^{H})}&=\sum_{k=1}^{K}{tr(\textbf{V}^{H}\textbf{V}\textbf{D}_{k})}\\
&=\sum_{k=1}^{K}{tr(\textbf{D}_{k})}=\sum_{k=1}^{K}{P_{k}}\leq{P},\\
\end{aligned}
\end{eqnarray}
\begin{eqnarray}
\begin{aligned}
\text{MAC:}\quad \sum_{k=1}^{K}{tr(\textbf{W}_{k}\textbf{Q}_{k}\textbf{W}^{H}_{k})}&=\sum_{k=1}^{K}{tr(\textbf{W}^{H}_{k}\textbf{W}_{k}\textbf{Q}_{k})}\\
&=\sum_{k=1}^{K}{tr(\textbf{Q}_{k})}=\sum_{k=1}^{K}{P_{k}}\leq{P},
\end{aligned}
\end{eqnarray}
\end{subequations}
which are met when both $\textbf{V}$ and $\textbf{W}_{k}$ be semi-unitary. These constraint will be considered to find the optimum variables in next sections. Suppose that the reduced SVD of $\textbf{B}^{-1/2}_{k}\tilde{\textbf{H}}^{H}_{k}\textbf{A}^{-1/2}_{k}$ with size $N_{s}\times{N_{s}}$ and rank $r$ is as below
\begin{equation} \label{eq.14}
\textbf{B}^{-1/2}_{k}\tilde{\textbf{H}}^{H}_{k}\textbf{A}^{-1/2}_{k}=\textbf{F}_{k}\boldsymbol{\Lambda}_{k}\textbf{G}^{H}_{k},
\end{equation}
where $\textbf{F}_{k}$ is the $N_{s}\times{r}$ semi-unitary matrix that contains the left orthonormal singular vectors of $\textbf{B}^{-1/2}_{k}\tilde{\textbf{H}}^{H}_{k}\textbf{A}^{-1/2}_{k}$ such that $\textbf{F}^{H}_{k}\textbf{F}_{k}=\textbf{I}_{r}$,  $\boldsymbol{\Lambda}_{k}$ is $r\times{r}$ diagonal matrix contains the singular values and $\textbf{G}_{k}$ is the $N_{s}\times{r}$ semi-unitary matrix corresponding to the right orthonormal singular vectors of $\textbf{B}^{-1/2}_{k}\tilde{\textbf{H}}^{H}_{k}\textbf{A}^{-1/2}_{k}$ such that $\textbf{G}^{H}_{k}\textbf{G}_{k}=\textbf{I}_{r}$. It is worth mentioning that the precoding matrix $V$ is semi-unitary and thus the first term of $B_j$ in (10) is always positive definite. Additionally, the second term of $B_j$ is of the Hermitian form and is always positive semi-definite (independent of the ranks of $H_k$ , $Q_k$ and $W_k$), the sum of the first term and the second will be always positive definite and consequently $B_j$ will be invertible. To transfer the optimum covariance matrices from MAC to BC to satisfy $R^{BC}_{j}=R^{MAC}_{j}$ for any $j\in{[1,2,...,K]}$, $\textbf{D}_{k}$ can be derived by below transformations \cite{r28}
\begin{equation} \label{eq.15}
\textbf{D}_{k}=\textbf{B}^{-1/2}_{k}\textbf{F}_{k}\textbf{G}^{H}_{k}\textbf{A}^{+1/2}_{k}\textbf{Q}_{k}\textbf{A}^{+1/2}_{k}\textbf{G}_{k}\textbf{F}^{H}_{k}\textbf{B}^{-1/2}_{k},
\end{equation}
It has been demonstrated in \cite{r28} that by the transformation of (15), $R^{BC}_{j}=R^{MAC}_{j}$ will be met. Then, the main concern is to find the optimum $\textbf{Q}_{k}$ in (12b) and transform it to BC by (15) to derive the same optimal sum rate in (12a).

In the next section, first we will focus on solving the problem (12b). Then after finding optimum precoding and combining and transmit covariance matrices, the duality theorem will be used to find the optimum dual covariance matrices in BC.

\section{Optimum hybrid beamforming and covariance matrices}\label{Optimum hybrid}

As discussed in previous sections, under the same total power constraint and with channel reciprocity assumption, the sum rate capacity region of the BC and MAC are equal. Thus, we consider the sum rate capacity of the MAC which is concave and then will transform the solutions to the dual BC.

\subsection{Optimum hybrid beamforming and covariances in MAC} \label{optimum mac}

Consider the sum rate maximization problem of MIMO MAC in (12b). The cost function can be simplified as

\begin{equation} \label{eq.16}
\begin{split}
&R^{MAC}=\frac{1}{2}\log{\big|\textbf{V}^{H}\textbf{V}+\textbf{V}^{H}\big(\sum_{k=1}^{K}{\textbf{H}^{H}_{k}\textbf{W}_{k}\textbf{Q}_{k}\textbf{W}^{H}_{k}\textbf{H}_{k}}\big)\textbf{V}\big|}\\
&-\frac{1}{2}\log{\big|\textbf{V}^{H}\textbf{V}\big|}\\
&=\frac{1}{2}\log{\frac{|\textbf{V}^{H}\textbf{V}+\textbf{V}^{H}\big(\sum_{k=1}^{K}{\textbf{H}^{H}_{k}\textbf{W}_{k}\textbf{Q}_{k}\textbf{W}^{H}_{k}\textbf{H}_{k}}\big)\textbf{V}|}{|\textbf{V}^{H}\textbf{V}|}}\\
&=\frac{1}{2}\log{\big|\textbf{I}_{N_{s}}+(\textbf{V}^{H}\textbf{V})^{-1}\textbf{V}^{H}\big(\sum_{k=1}^{K}{\textbf{H}^{H}_{k}\textbf{W}_{k}\textbf{Q}_{k}\textbf{W}^{H}_{k}\textbf{H}_{k}}\big)\textbf{V}}\big|,
\end{split}
\end{equation}
Suppose that the reduced SVD of $\textbf{V}^{H}\textbf{V}=\textbf{S}_{v}\boldsymbol{\Sigma}_{v}\textbf{S}^{H}_{v}$, where $\textbf{S}_{v}$ is the $N_{s}\times{N_{s}}$ unitary matrix which its columns are the left singular vectors of $\textbf{V}^{H}\textbf{V}$ and $\boldsymbol{\Sigma}_{v}$ is a $N_{s}\times{N_{s}}$ diagonal matrix contains the singular values of $\textbf{V}^{H}\textbf{V}$. Then the term $(\textbf{V}^{H}\textbf{V})^{-1}$ in (16) can be rewritten as $(\textbf{V}^{H}\textbf{V})^{-1}=\textbf{S}_{v}\boldsymbol{\Sigma}^{-1}_{v}\textbf{S}^{H}_{v}=\textbf{S}_{v}\boldsymbol{\Sigma}^{-1/2}_{v}\boldsymbol{\Sigma}^{-1/2}_{v}\textbf{S}^{H}_{v}$. Therefore, (16) can be expressed as
\begin{equation} \label{eq.17}
\begin{split}
R^{MAC}=\frac{1}{2}\text{log}\bigg|\textbf{I}_{N_{s}}+&(\textbf{S}_{v}\boldsymbol{\Sigma}^{-1/2}_{v})(\boldsymbol{\Sigma}^{-1/2}_{v}\textbf{S}^{H}_{v})\times\bigg(\\
&\textbf{V}^{H}\big(\sum_{k=1}^{K}{\textbf{H}^{H}_{k}\textbf{W}_{k}\textbf{Q}_{k}\textbf{W}^{H}_{k}\textbf{H}_{k}}\big)\textbf{V}\bigg)\bigg|,
\end{split}
\end{equation}
Using the fact that $\log{\big|\textbf{I}_{m}+\textbf{Z}_{1}\textbf{Z}_{2}\big|}=\log{\big|\textbf{I}_{n}+\textbf{Z}_{2}\textbf{Z}_{1}\big|}$ for any $\textbf{Z}_{1}\in{\mathbb{C}^{m\times{n}}}$ and $\textbf{Z}_{2}\in{\mathbb{C}^{n\times{m}}}$, (17) can be derived as
\begin{equation} \label{eq.18}
\begin{split}
R^{MAC}=\frac{1}{2}\log{\bigg|\textbf{I}_{N_{s}}+\tilde{\textbf{V}}^{H}\bigg(\sum_{k=1}^{K}{\textbf{H}^{H}_{k}\textbf{W}_{k}\textbf{Q}_{k}\textbf{W}^{H}_{k}\textbf{H}_{k}}\bigg)\tilde{\textbf{V}}\bigg|},
\end{split}
\end{equation}
where $\tilde{\textbf{V}}\triangleq\textbf{V}\textbf{S}_{v}\boldsymbol{\Sigma}^{-1/2}_{v}$ is the $N\times{N_{s}}$ semi-unitary matrix because
\begin{subequations}\label{eq.19}
\begin{eqnarray}
\begin{aligned} \tilde{\textbf{V}}^{H}\tilde{\textbf{V}}&=\boldsymbol{\Sigma}^{-1/2}_{v}\textbf{S}^{H}_{v}(\textbf{V}^{H}\textbf{V})\textbf{S}_{v}\boldsymbol{\Sigma}^{-1/2}_{v}\\
&=\boldsymbol{\Sigma}^{-1/2}_{v}\textbf{S}^{H}_{v}(\textbf{S}_{v}\boldsymbol{\Sigma}_{v}\textbf{S}^{H}_{v})\textbf{S}_{v}\boldsymbol{\Sigma}^{-1/2}_{v}\\
&=\boldsymbol{\Sigma}^{-1/2}_{v}\boldsymbol{\Sigma}_{v}\boldsymbol{\Sigma}^{-1/2}_{v}=\textbf{I}_{N_{s}},
\end{aligned}
\end{eqnarray}
\begin{eqnarray}
\begin{aligned}
\tilde{\textbf{V}}\tilde{\textbf{V}}^{H}&=\textbf{V}\textbf{S}_{v}\boldsymbol{\Sigma}^{-1/2}_{v}\boldsymbol{\Sigma}^{-1/2}_{v}\textbf{S}^{H}_{v}\textbf{V}^{H}\\
&=\textbf{V}\textbf{S}_{v}\boldsymbol{\Sigma}^{-1}_{v}\textbf{S}^{H}_{v}\textbf{V}^{H}\\
&=\textbf{V}(\textbf{V}^{H}\textbf{V})^{-1}\textbf{V}^{H},
\end{aligned}
\end{eqnarray}
\end{subequations}
The optimization problem of (12b) related to the MAC can be rewritten as
\begin{equation}\label{eq.20}
\begin{aligned}
&\max_{\textbf{V},\textbf{W}_{k},\textbf{Q}_{k}} \quad \frac{1}{2}\log{\bigg|\textbf{I}_{N_{s}}+\tilde{\textbf{V}}^{H}\bigg(\sum_{k=1}^{K}{\textbf{H}^{H}_{k}\textbf{W}_{k}\textbf{Q}_{k}\textbf{W}^{H}_{k}\textbf{H}_{k}}\bigg)\tilde{\textbf{V}}\bigg|}\\
&\textrm{s.t.} \quad \begin{cases}
\sum_{k=1}^{K}{tr(\textbf{W}_{k}\textbf{Q}_{k}\textbf{W}^{H}_{k})}\leq{P},\\
\big|\textbf{W}_{RF,k}(i,j)\big|=1,
\end{cases}
\end{aligned}
\end{equation}
The cost function in (20) is neither concave nor convex with respect to the variables $\textbf{V}$, $\textbf{W}_{k}$ and $\textbf{Q}_{k}$. Here, we will use an alternating method to solve above problem in three steps. In each step, it is assumed that two variables are constant and the optimum third variable will be found. This iteration will be continued till meet the final criterion. It must be noted that with this assumption, the problem in each step will be concave with respect to one set of variable($\textbf{V}$, $\textbf{W}_{k}$ or $\textbf{Q}_{k}$ respectively). Simulation results demonstrate fast convergence trend of the proposed algorithm compared to other numerical algorithms. The details of the proposed algorithm is as follow.

\emph{step1:} In the first step, it is assumed that $\textbf{V}$ and $\textbf{W}_{k}$ are constant for all $k\in{1,2,\ldots,K}$, then the iterative waterfilling algorithm as in \cite{r29} will be used to find the optimum individual covariance matrices. In a $K$-user MAC, $\textbf{Q}_{k}$ is an optimum solution to (20) if and only if $\textbf{Q}_{k}$ be the single-user waterfilling covariance matrix of the channel $\textbf{W}^{H}_{k}\textbf{H}_{k}\tilde{\textbf{V}}$ with $\textbf{I}_{N_{s}}+\tilde{\textbf{V}}^{H}\big(\sum_{j\neq{k}}{\textbf{H}^{H}_{j}\textbf{W}_{j}\textbf{Q}_{j}\textbf{W}^{H}_{j}\textbf{H}_{j}}\big)\tilde{\textbf{V}}$ as the covariance of the noise term, for all $k\in{1,2,\ldots,K}$. For moderate or high signal to noise ratios (SNRs), the derived $\textbf{Q}_{k}$s are full rank. For low SNR regimes, $\textbf{Q}_{k}$ may be rank deficient and this will affect the number of data streams corresponding to each user. In next steps, this will be discussed.  As demonstrated in \cite{r29}, the iterative waterfilling algorithm will be converged to the optimum point very fast.

\emph{step2:} In this step, suppose $\textbf{W}_{k}$s are constant and $\textbf{Q}_{k}$s are constant too as found in the first step. To find optimum $\tilde{\textbf{V}}$ or $\textbf{V}$, assume that $\sum_{k=1}^{K}{\textbf{H}^{H}_{k}\textbf{W}_{k}\textbf{Q}_{k}\textbf{W}^{H}_{k}\textbf{H}_{k}}=\textbf{A}^{H}\textbf{A}$ where $\textbf{A}\in{\mathbb{C}^{r_{t}\times{N}}}$ and $r_{t}=rank(A)=rank(\sum_{k=1}^{K}{\textbf{H}^{H}_{k}\textbf{W}_{k}\textbf{Q}_{k}\textbf{W}^{H}_{k}\textbf{H}_{k}})$. Thus, (18) can be rewritten as
\begin{equation} \label{eq.21}
\begin{split}
R^{MAC}&=\frac{1}{2}\log{\big|\textbf{I}_{N_{s}}+\tilde{\textbf{V}}^{H}\textbf{A}^{H}\textbf{A}\tilde{\textbf{V}}\big|}\\
&=\frac{1}{2}\log{\big|\textbf{I}_{r_{t}}+\textbf{A}\tilde{\textbf{V}}\tilde{\textbf{V}}^{H}\textbf{A}^{H}\big|},
\end{split}
\end{equation}
Note that the power constraint should be considered as $tr(\tilde{\textbf{V}}\tilde{\textbf{V}}^{H})=tr(\tilde{\textbf{V}}^{H}\tilde{\textbf{V}})=tr(\textbf{I}_{N_{s}})=N_{s}$. Maximization of (21) is like to find the optimum matrix $\tilde{\textbf{V}}\tilde{\textbf{V}}^{H}$ in a downlink of a single user MIMO with channel matrix $\textbf{A}$. For moderate and high SNR regimes, $\textbf{Q}_{k}$s are full rank and $r_{t}=Kd=N_{s}$, and for low SNR values, $\textbf{Q}_{k}$s may be rank deficient and $r_{t}=rank(\sum_{k=1}^{K}{\textbf{H}^{H}_{k}\textbf{W}_{k}\textbf{Q}_{k}\textbf{W}^{H}_{k}\textbf{H}_{k}})\leq{Kd}=N_{s}$. Here without loss of generality, we assumed that $\textbf{A}$ is full rank and $rank(\textbf{A})=\min{(r_{t},N)}=r_{t}$; although we can assume $N_{s}$ number of rows for $\textbf{A}$ with rank $r_{t}$ such that $r_{t}\leq{N_{s}}$ and $\sum_{k=1}^{K}{\textbf{H}^{H}_{k}\textbf{W}_{k}\textbf{Q}_{k}\textbf{W}^{H}_{k}\textbf{H}_{k}}=\textbf{A}^{H}\textbf{A}$. Then $\textbf{A}$ can be derived easily using the reduced SVD of $\sum_{k=1}^{K}{\textbf{H}^{H}_{k}\textbf{W}_{k}\textbf{Q}_{k}\textbf{W}^{H}_{k}\textbf{H}_{k}}$. Suppose that the reduced SVD decomposition of $\textbf{A}$ as $\textbf{A}=\textbf{F}_{a}\Sigma_{a}\textbf{G}^{H}_{a}$ where $\textbf{F}_{a}$ is the $r_{t}\times{r_{t}}$ unitary matrix, $\boldsymbol{\Sigma}_{a}$ is the $r_{t}\times{r_{t}}$ diagonal matrix contains the singular values of $\textbf{A}$ and $\textbf{G}_{a}$  is the $N\times{r_{t}}$ semi-unitary matrix (i.e., $\textbf{G}^{H}_{a}\textbf{G}_{a}=\textbf{I}_{r_{t}}$). By defining $\tilde{\textbf{V}}\tilde{\textbf{V}}^{H}\triangleq{\textbf{T}_{v}}$, from (21) we have
\begin{equation} \label{eq.22}
\begin{split}
R^{MAC}&=\frac{1}{2}\log{\big|\textbf{I}_{r_{t}}+\textbf{A}\tilde{\textbf{V}}\tilde{\textbf{V}}^{H}\textbf{A}^{H}\big|}\\
&=\frac{1}{2}\log{\big|\textbf{I}_{r_{t}}+\textbf{F}_{a}\Sigma_{a}\textbf{G}^{H}_{a}\textbf{T}_{v}\textbf{G}_{a}\Sigma^{H}_{a}\textbf{F}^{H}_{a}\big|}\\
&=\frac{1}{2}\log{\big|\textbf{I}_{r_{t}}+\textbf{F}^{H}_{a}\textbf{F}_{a}\Sigma_{a}\textbf{G}^{H}_{a}\textbf{T}_{v}\textbf{G}_{a}\Sigma^{H}_{a}\big|}\\
&=\frac{1}{2}\log{\big|\textbf{I}_{r_{t}}+\Sigma_{a}\textbf{G}^{H}_{a}\textbf{T}_{v}\textbf{G}_{a}\Sigma^{H}_{a}\big|}\\
&=\frac{1}{2}\log{\big|\textbf{I}_{r_{t}}+\Sigma_{a}\tilde{\textbf{T}}_{v}\Sigma^{H}_{a}\big|},\\
\end{split}
\end{equation}
where in the last equation $\tilde{\textbf{T}}_{v}$ with size $r_{t}\times{r_{t}}$ is defined as $\tilde{\textbf{T}}_{v}\triangleq{\textbf{G}^{H}_{a}\textbf{T}_{v}\textbf{G}_{a}}$. According to the well-known waterfilling algorithm, the optimum $\tilde{\textbf{T}}_{v}$ to maximize (22) is derived as the diagonal matrix. Next, we have to find $\textbf{T}_{v}$ and $\tilde{\textbf{V}}$ accordingly. By using semi-unitary property of $\textbf{G}_{a}$ and multiplying each side of the equation $\tilde{\textbf{T}}_{v}\triangleq{\textbf{G}^{H}_{a}\textbf{T}_{v}\textbf{G}_{a}}$ by $\textbf{G}_{a}$ from left and $\textbf{G}^{H}_{a}$ from right, $\textbf{T}_{v}$ will be achieved as
\begin{equation} \label{eq.23}
\begin{split}
&\textbf{G}_{a}\textbf{G}^{H}_{a}\textbf{T}_{v}\textbf{G}_{a}\textbf{G}^{H}_{a}=\textbf{G}_{a}\tilde{\textbf{T}}_{v}\textbf{G}^{H}_{a}\\
&\Rightarrow\textbf{T}_{v}=(\textbf{G}_{a}\textbf{G}^{H}_{a})^{\dag}\textbf{G}_{a}\tilde{\textbf{T}}_{v}\textbf{G}^{H}_{a}(\textbf{G}_{a}\textbf{G}^{H}_{a})^{\dag},
\end{split}
\end{equation}
As defined before, $\textbf{T}_{v}\triangleq{\tilde{\textbf{V}}\tilde{\textbf{V}}^{H}}$, thus $\tilde{\textbf{V}}$ will be derived as
\begin{equation} \label{eq.24}
\tilde{\textbf{V}}=(\textbf{G}_{a}\textbf{G}^{H}_{a})^{\dag}\textbf{G}_{a}\tilde{\textbf{T}}^{1/2}_{v},
\end{equation}
From (24) we have
\begin{equation} \label{eq.25}
\begin{split}
\tilde{\textbf{V}}^{H}\tilde{\textbf{V}}&=\tilde{\textbf{T}}^{1/2}_{v}\textbf{G}^{H}_{a}((\textbf{G}_{a}\textbf{G}^{H}_{a})^{\dag})^{H}(\textbf{G}_{a}\textbf{G}^{H}_{a})^{\dag}\textbf{G}_{a}\tilde{\textbf{T}}^{1/2}_{v}\\
&=\tilde{\textbf{T}}^{1/2}_{v}\tilde{\textbf{T}}^{1/2}_{v}=\tilde{\textbf{T}}_{v},
\end{split}
\end{equation}
Since the matrix $\tilde{\textbf{V}}$ is semi-unitary ($\tilde{\textbf{V}}^{H}\tilde{\textbf{V}}=\textbf{I}_{r_{t}}$), it is concluded that $\tilde{\textbf{T}}_{v}=\textbf{I}_{r_{t}}$). According to (19), $\tilde{\textbf{V}}\tilde{\textbf{V}}^{H}=\textbf{V}(\textbf{V}^{H}\textbf{V})^{-1}\textbf{V}^{H}$ and thus $\tilde{\textbf{V}}=\textbf{V}(\textbf{V}^{H}\textbf{V})^{-1/2}$. One of the possible solutions to find $\textbf{V}$ is $\textbf{V}=\tilde{\textbf{V}}$. Finally,
\begin{equation} \label{eq.26}
\textbf{V}=(\textbf{G}_{a}\textbf{G}^{H}_{a})^{\dag}\textbf{G}_{a},
\end{equation}
which is a semi-unitary matrix. Therefore, in the dual BC, it will not change the power of transmitted signals which is matched with the power constraint, too.

\emph{step3:} In the last step, suppose that $\textbf{Q}_{k}$s and $\textbf{V}$ are constant as derived in the previous two steps and we are focusing to find optimum $\textbf{W}_{k}$s. The sum rate expression in (18) can be represented as
\begin{equation} \label{eq.27}
\begin{split}
R^{MAC}&=\frac{1}{2}\log{\bigg|\textbf{I}_{N_{s}}+\tilde{\textbf{V}}^{H}\bigg(\sum_{k=1}^{K}{\textbf{H}^{H}_{k}\textbf{W}_{k}\textbf{Q}_{k}\textbf{W}^{H}_{k}\textbf{H}_{k}}\bigg)\tilde{\textbf{V}}\bigg|}\\
&=\frac{1}{2}\log{\bigg|\textbf{I}_{N_{s}}+\bigg(\sum_{k=1}^{K}{\tilde{\textbf{V}}^{H}\textbf{H}^{H}_{k}\textbf{W}_{k}\textbf{Q}_{k}\textbf{W}^{H}_{k}\textbf{H}_{k}\tilde{\textbf{V}}}\bigg)\bigg|}\\
&=\frac{1}{2}\log{\bigg|\textbf{I}_{N_{s}}+\bigg(\sum_{k=1}^{K}{\textbf{C}^{H}_{k}\textbf{W}_{k}\textbf{Q}_{k}\textbf{W}^{H}_{k}\textbf{C}_{k}}\bigg)\bigg|}\\
&=\frac{1}{2}\log{\bigg|\textbf{I}_{N_{s}}+\bigg(\sum_{k=1}^{K}{\textbf{C}^{H}_{k}\tilde{\textbf{Q}}_{k}\textbf{C}_{k}}\bigg)\bigg|},\\
\end{split}
\end{equation}
Where $\textbf{C}_{k}\triangleq{\textbf{H}_{k}\textbf{V}}$ and $\tilde{\textbf{Q}}_{k}\in{\mathbb{C}^{M\times{M}}}$ is defined as the modified covariance matrix as $\tilde{\textbf{Q}}_{k}\triangleq{\textbf{W}_{k}\textbf{Q}_{k}\textbf{W}^{H}_{k}}$ with reduced rank, i.e, $rank(\tilde{\textbf{Q}}_{k})=d_{k}<M$. Note that to find the optimum $\textbf{W}_{k}$ and to maintain the same power level in both MAC and BC to employ the duality theorem, the power constraint of the MAC must be considered as $tr(\tilde{\textbf{Q}}_{k})=tr(\textbf{W}_{k}\textbf{Q}_{k}\textbf{W}^{H}_{k})=tr(\textbf{W}^{H}_{k}\textbf{W}_{k}\textbf{Q}_{k})=tr(\textbf{Q}_{k})=P_{k}$ which required semi-unitary property of $\textbf{W}_{k}$, i.e., $\textbf{W}^{H}_{k}\textbf{W}_{k}=\textbf{I}_{d}$. Next, to find the optimum $\textbf{W}_{k}$, we have to find the optimum modified covariance matrix first, so we have
\begin{equation}\label{eq.28}
\begin{aligned}
&\max_{\tilde{\textbf{Q}}_{k}, k\in{\{1,2,...,K\}}} \quad \frac{1}{2}\log{\bigg|\textbf{I}_{N_{s}}+\bigg(\sum_{k=1}^{K}{\textbf{C}^{H}_{k}\tilde{\textbf{Q}}_{k}\textbf{C}_{k}}\bigg)\bigg|}\\
&\textrm{s.t.} \quad \sum_{k=1}^{K}{tr(\tilde{\textbf{Q}}_{k})}=\sum_{k=1}^{K}{P_{k}}\leq{P},\\
\end{aligned}
\end{equation}
Above maximization problem is same as sum rate maximization problem in MU-MIMO MAC with $\textbf{C}_{k}\in{\mathbb{C}^{M\times{N_{s}}}}$ as the effective channel and $\textbf{I}_{N_{s}}$ as the covariance matrix of the effective noise. Thus, $\tilde{\textbf{Q}}_{k}$ can be found using iterative waterfilling algorithm \cite{r28} considering that the assigned power to $M-d_{k}$ subchannels out of M subchannels should be zero. In other words, the power should be assigned to the $d_{k}$ largest Eigen values of the channel covariance to meet the reduced rank constraint. After finding $\tilde{\textbf{Q}}_{k}$, suppose that its reduced SVD decomposition as $\tilde{\textbf{Q}}_{k}=\tilde{\textbf{F}}_{qk}\tilde{\boldsymbol{\Sigma}}_{qk}\tilde{\textbf{F}}^{H}_{qk}$, where $\tilde{\textbf{F}}_{qk}\in{\mathbb{C}^{M\times{d_{k}}}}$ is the semi-unitary matrix and $\tilde{\boldsymbol{\Sigma}}_{qk}$ is the diagonal matrix contains the non-zero singular values of $\tilde{\textbf{Q}}_{k}$. Then  $\textbf{W}_{k}$ will be derived as
\begin{equation} \label{eq.29}
\begin{split}
\tilde{\textbf{Q}}_{k}&=\textbf{W}_{k}\textbf{Q}_{k}\textbf{W}^{H}_{k}=\textbf{W}_{k}\textbf{Q}^{1/2}_{k}\textbf{Q}^{1/2}_{k}\textbf{W}^{H}_{k}\\
&\Rightarrow \tilde{\textbf{F}}_{qk}\tilde{\boldsymbol{\Sigma}}^{1/2}_{qk}=\textbf{W}_{k}\textbf{Q}^{1/2}_{k}\\
&\Rightarrow \textbf{W}_{k}=\tilde{\textbf{F}}_{qk}\tilde{\boldsymbol{\Sigma}}^{1/2}_{qk}\textbf{Q}^{-1/2}_{k},\\
\end{split}
\end{equation}
These three steps will be continued till converging to the optimum solutions. After finding the overall beamforming matrices, the digital and analog hybrid beamformers can be found by solving $V=V_{RF} V_D$ where the components in $V_{RF}$ are constant modulus ($V_{RF}=e^{\theta_{i,j}}$). As demonstrated in [7], to find the solution, it is sufficient that the number of RF chains in hybrid structure be greater than or equal to twice the number of data streams. This will guarantee the existence of the solutions for analog and digital beamforming matrices. It must be noted that the same is true for $W_k=W_{RF,k} W_{D,k}$ to achieve the individual analog and digital combiners per user. Additionally, it is assumed that infinite resolution phase shifters are available to produce analog beamformers and combiners. In practical cases, infinite alphabet can be assumed and then the optimum solution of phase shifters can be mapped on to available finite phase shifters which is the least square of optimum analog beamformer minus the available analog beamformer.

\emph{Corollary 1}: There are three set of variables in the proposed algorithm ($\textbf{V}$, $\textbf{W}_{k}$ and $\textbf{Q}_{k}$). In each step, two sets are assumed to be fixed and the third set is optimized. After the first iteration, the achieved variables will be the initial values for the next iteration. Thus, in each element update step of the proposed algorithm, the objective function of (20) increases (or at least does not decrease); consequently the convergence of the algorithm is guaranteed.

\subsection{Optimum hybrid beamforming and covariances in BC} \label{optimum BC}

After finding the optimum individual user covariance matrices and hybrid digital and analog precoder and combiners in the MAC, MAC-BC duality will be employed to find the optimum transmit correlations in the BC. As discussed in previous sections, to maintain the channel reciprocity and to have same transmitted power in both MAC and BC (to validate the MAC-BC duality theorem), we will assume both hybrid precoders and combiners must be semi-unitary. The optimum precoders and combiners in the MAC will be used in the BC too. Thus, the effective channel in the uplink will be Hermitian of the channel in the downlink. Finally, the individual covariance matrices in the BC can be found by using MAC-BC transformation in (15).

\begin{figure*}[ht]
\centering
\begin{minipage}[b]{0.45\linewidth}
\includegraphics[width=\textwidth]{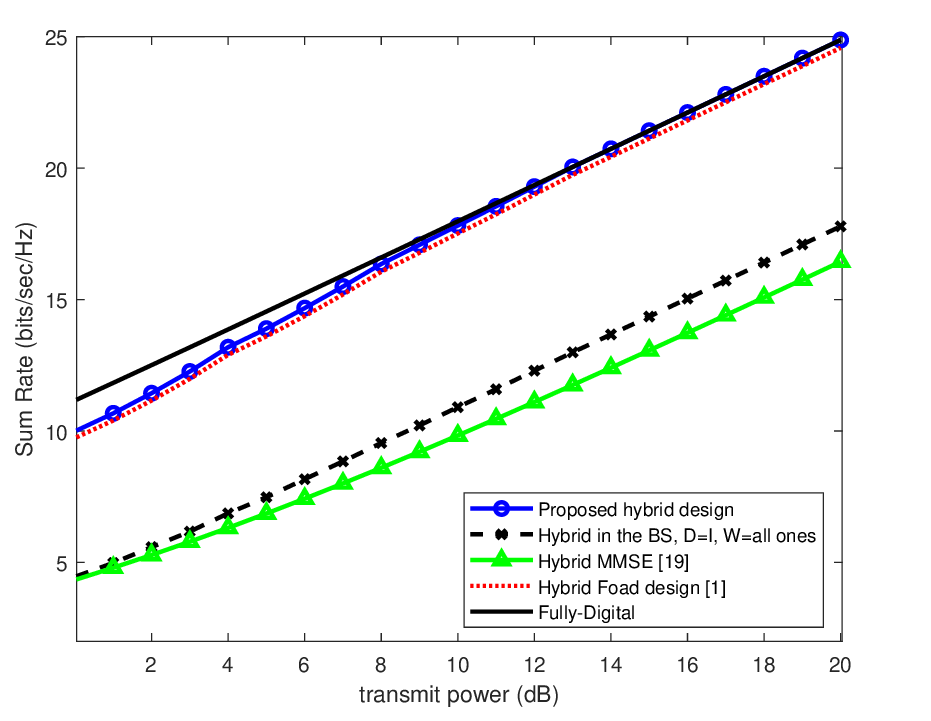}
\caption{Sum rate achieved by different methods in point-to-point MIMO where $N=64$, $M=3$, and $d=2$}
\label{fig2}
\end{minipage}
\quad
\begin{minipage}[b]{0.45\linewidth}
\includegraphics[width=\textwidth]{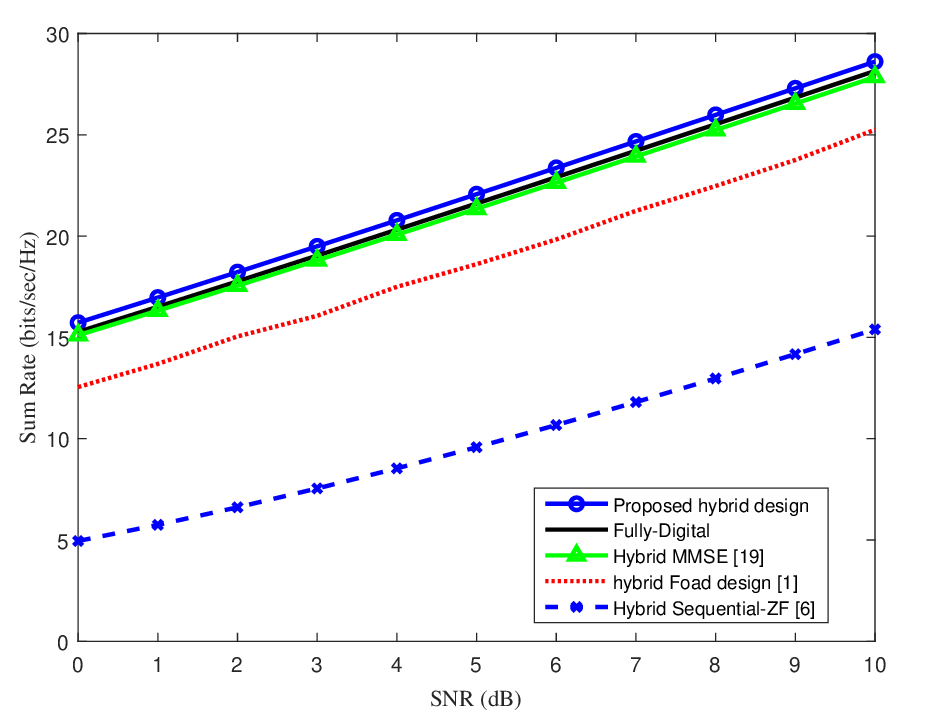}
\caption{Sum rate achieved by different methods in MU-MISO where $K=8$, $N=64$, $M=1$, and $d=1$}
\label{fig3}
\end{minipage}
\end{figure*}

\section{Simulation Results} \label{sec:Analysis}

In this section, simulation results are presented to show the performance of the proposed algorithm for single user point-to-point MIMO, MU-MISO and MU-MIMO systems, respectively and to compare them with the existing hybrid designs. In the following simulations, an environment channel with $L=15$ scatterers between each individual user and the BS is assumed that has random angles of arrival and departure with uniform distribution between $0$ and $2\pi$. Furthermore, half wavelength spacing between antenna elements in both BS and user equipment is assumed. For each scenario, the average sum rate is plotted versus SNR (as represented in prior sections and without loss of generality, it is assumed that the noise power is unity and thus $\text{SNR}=P=\sum_{k}{P_{k}}$). To utilize the minimum number of analog RF chains, $N^{RF}_{t}=2N_{s}=2Kd$ is assumed in the BS to guarantee the existence of the solutions for digital and analog precoders. In addition, the same is true for the number of analog RF chains in the user side. All the key simulation parameters are summarized in Table II.

\begin{figure*}[ht]
\centering
\begin{minipage}[b]{0.45\linewidth}
\includegraphics[width=\textwidth]{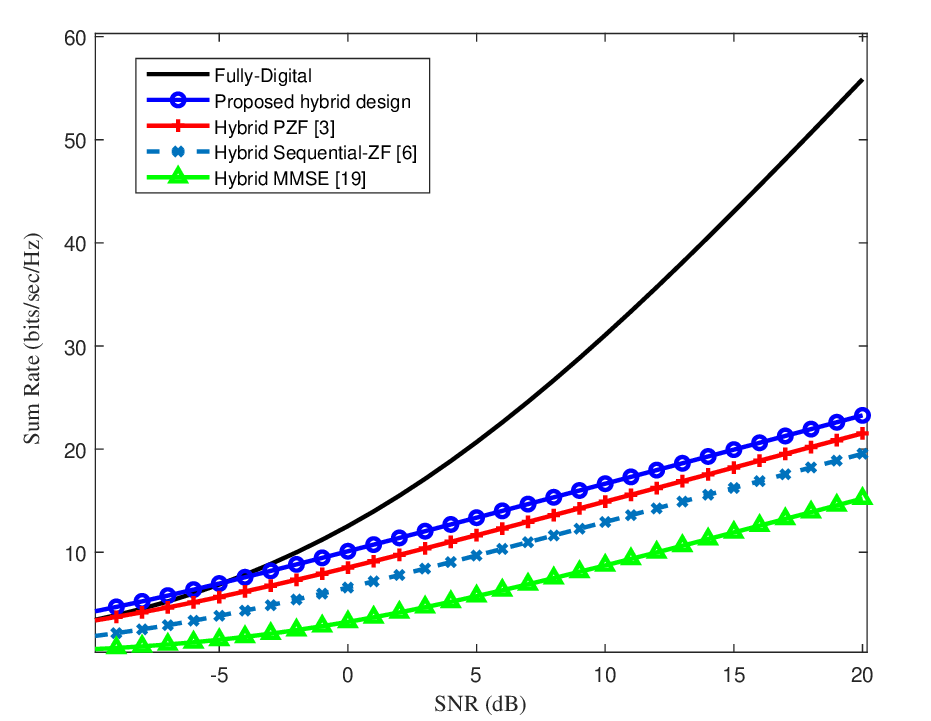}
\caption{Sum rate achieved by different methods in MU-MIMO where $K=4$, $N=32$, $M=4$, and $d=1$}
\label{fig4}
\end{minipage}
\quad
\begin{minipage}[b]{0.45\linewidth}
\includegraphics[width=\textwidth]{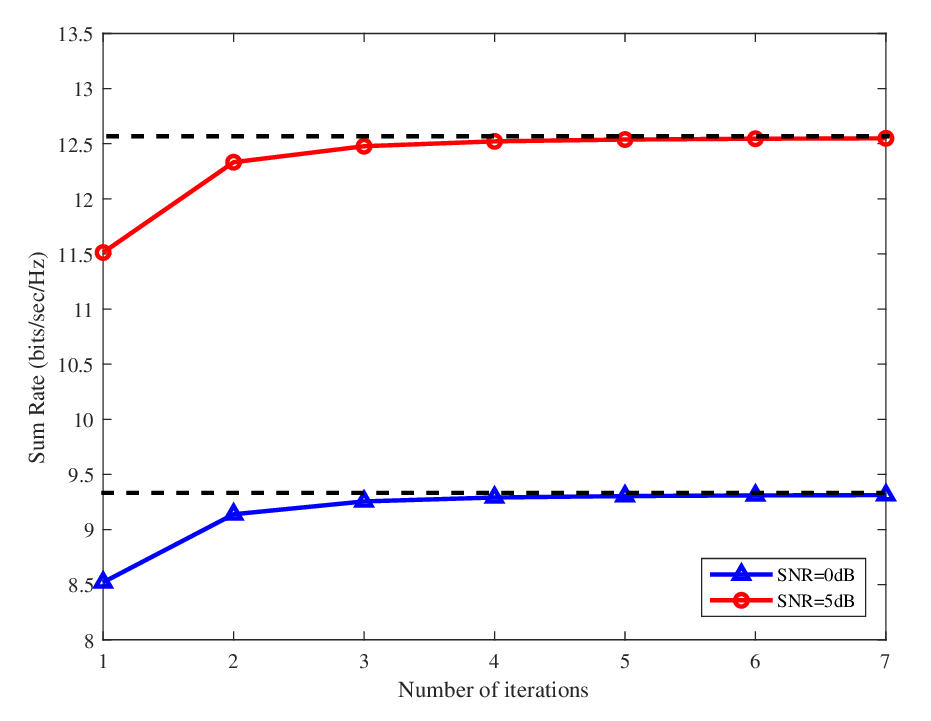}
\caption{Convergence trend of the sum rate capacity in proposed method for $K=4$, $N=32$, $M=4$, and $d=1$}
\label{fig5}
\end{minipage}
\end{figure*}

\begin{table}
\centering
\caption{Simulation parameters}
    \begin{tabular}{ | c | c | c | c | c |}
    \hline
    Figure number & $K$ & $N$ & $M$ & $d$\\ \hline
    Fig. 2: Point-to-point MIMO & 1 & 64 & 3 & 2 \\ \hline
    Fig. 3: MU-MISO & 8 & 64 & 1 & 1 \\ \hline
    Fig. 4: MU-MIMO & 4 & 32 & 4 & 1 \\
    \hline
    \end{tabular}
\end{table}

\subsection{Performance analysis of hybrid point-to-point MIMO} \label{P2P MIMO}

In the first simulation, we consider a single-user MIMO system with $M=3$ antennas at the user equipment and $N=64$ antennas in the BS which transmits $d=2$ data streams per user. Here, traditional schemes including MMSE in \cite{r10} and iterative hybrid beamforming in \cite{r7} are simulated. Also hybrid beamforming design with identity covariance matrix is simulated. As shown in Fig. 2, the proposed design outperforms other traditional schemes in terms of sum rate with much lower complexity. It must be noted that the proposed design is the capacity achieving method among all hybrid designs with pre-assumed variables and thus is optimum, how ever in high SNR values, the performance of the proposed method is matched with fully-digital scheme, too.

\subsection{Performance analysis of hybrid MU-MISO} \label{MU-MISO MIMO}

In the second simulation, a MU-MISO system with $K=8$ users that each of them employs single antenna is considered. Suppose that the BS employs $N=64$ antennas that transmits $d=1$ data stream. Sum rate of the proposed scheme is illustrated in Fig. 3, and is compared with hybrid beamforming design in \cite{r7}, hybrid sequential-ZF \cite{r22} and the hybrid MMSE scheme \cite{r10}. As shown in this figure, the proposed design performs much better than other schemes in terms of the output sum rate. The proposed design which is the sum rate capacity achieving method outperforms the BD (fully-digital block diagonalization scheme) in this scenario. In this example, the proposed scheme with $N_{s}=8$ digital RF chains performs about $0.4$ dB better than fully-digital BD scheme which requires $N=64$ digital RF chains. Although there is still a gap between capacity of hybrid structure and non-hybrid fully digital scheme that is inevitable due to the imperfect approximation of the precoder matrices, this gap can be reduced by increasing the number of iterations and employing high resolution phase shifters in the proposed method. Hence, to reduce the gap, we can use more expensive high resolution phase shifters to approach the optimum condition as much as possible.

\subsection{Performance analysis of hybrid MU-MIMO} \label{MU-MIMO MIMO}

In this part of simulation, consider a MU-MIMO system with $N=32$ antennas at the BS that serves $K=4$ users each equipped with $M=4$ antennas. It is assumed that $d=1$ data stream per user is being transmitted. The achieved sum rate by the proposed design is compared with phased ZF (PZF) method \cite{r12}, the modified hybrid sequential-ZF \cite{r22} which zero forces the interference terms with lower complexity than PZF and the traditional MMSE method \cite{r10} as illustrated in Fig. 4. Same as previous scenarios, the proposed method is the capacity achieving scheme among hybrid MU-MIMO systems. In this example, the proposed method achieves about $2.5$dB better performance than the hybrid MMSE method. As shown in Fig. 5, the convergence trend of the proposed method is exponential. Thus, the proposed algorithm will be converged very fast by increasing the number of iterations.

\emph{Corollary 2}: Assuming that the number of antennas at both ends in the same range, i.e., $M = O(N)$, it can be shown that the complexity order of each step in the proposed algorithm will be $O(N^3)$; As the convergence rate is exponential, the overall complexity order will be $O(N^3 \log(1/\varepsilon))$ which $\varepsilon$ is the absolute distance between achieved sum rate and the optimum sum rate. On the other hand, the complexity of finding the optimal beamformers using the exhaustive search method is exponential, $O(N^2 2^N)$ \cite{r7}. Consequently , the complexity order of the proposed design is much lower than the traditional schemes.

\section{CONCLUSION} \label{sec:Conclusion}

In this paper, we present an improved design for hybrid beamforming in MU-MIMO systems. Unlike previous schemes that utilized an identity covariance matrix for each user, our approach incorporates individual covariance matrices. By leveraging the MAC-BC duality theorem, we are able to obtain optimal hybrid beamforming and transmit covariance matrices. This theorem allows us to achieve the same sum rates as in the MAC, while utilizing closed-form solutions with low complexity.
To evaluate the performance of our proposed scheme, we conducted simulations in three scenarios: point-to-point MIMO, MU-MISO, and MU-MIMO. The simulation results demonstrate that our scheme outperforms existing designs, with significantly lower complexity.
Overall, our enhanced hybrid beamforming scheme considering individual covariance matrices offers improved performance in MU-MIMO systems. This makes it a promising solution for future wireless communication networks. Finally, it would be intriguing to explore the potential application of the proposed scheme within holographic MIMO systems, which calls for future researches.

\bibliographystyle{IEEEtran}
\bibliography{ref}

\end{document}